\documentclass[]{spie}  %>>> use for US letter paper
\usepackage{amsmath,amsfonts,amssymb}
\usepackage{graphicx}
\usepackage[colorlinks=true, allcolors=blue]{hyperref}
\usepackage{tikz}
\usepackage{adjustbox}
\usepackage{xcolor}
\usepackage{subcaption}
\usepackage{url}

\usepackage{makecell}

\usepackage{array} % For 'p{}' column specifier
\usepackage{booktabs} % For better table lines
\usepackage{arydshln} % For dashed lines

\usetikzlibrary{math} %needed tikz library

\title{Thermal architecture for a cryogenic super-pressure balloon payload: design and development of the Taurus flight cryostat.}

\author[a]{Simon Tartakovsky}
\author[b]{Alexandre E. Adler}
\author[c]{Jason E. Austermann}
\author[a]{Steven J. Benton}
\author[d]{Rick Bihary}
\author[g,c]{Malcolm Durkin} % Hannes added, for contributions to the multiplexer
\author[c]{Shannon M. Duff}
\author[e]{Jeffrey P. Filippini}
\author[a]{Aurelien A. Fraisse}
\author[f]{Thomas J.L.J. Gascard}
\author[e]{Sho M. Gibbs}
\author[a]{Suren Gourapura}
\author[f, b]{Jon E. Gudmundsson}
\author[h]{John W. Hartley}
\author[c]{Johannes Hubmayr}
\author[a]{William C. Jones}
\author[h]{Steven Li}
\author[d]{Jared L. May}
\author[a]{Johanna M. Nagy}
\author[c,g]{Kate Okun}
\author[a]{Ivan L. Padilla}
\author[h]{L. Javier Romualdez}
\author[c]{Michael R. Vissers}

\affil[a]{Department of Physics, Princeton University, Jadwin Hall, Princeton, NJ 08544, USA}

\affil[b]{The Oskar Klein Centre, Department of Physics, Stockholm University, AlbaNova, SE-10691 Stockholm, Sweden}
\affil[c]{National Institute of Standards and Technology, 325 Broadway Mailcode 817.03, Boulder, CO 80305, USA}
\affil[d]{Department of Physics, Case Western Reserve University, 10900 Euclid Ave, Cleveland, OH 44106, USA}
\affil[e]{Department of Physics, University of Illinois Urbana-Champaign, 1110 W Green St, Urbana, IL 61801, USA}
\affil[f]{Science Institute, University of Iceland, 107 Reykjavik, Iceland}
\affil[g]{Department of Physics, University of Colorado Boulder, Boulder, Colorado, USA}
\affil[h]{StarSpec Technologies Inc., Unit C-5, 1600 Industrial Avenue, Cambridge, ON N3H 4W5, Canada}

\authorinfo{Further author information: (Send correspondence to S.T.)\\S.T.: E-mail: simont@princeton.edu}

\begin{document} 
\maketitle

\renewcommand{\arraystretch}{1.15}

\begin{abstract}
We describe the cryogenic system being developed for Taurus: a super-pressure balloon-borne microwave polarimeter scheduled to fly in 2027.  The Taurus cryogenic system consists of a 660\,L liquid helium cryostat which achieves a base temperature of $\leq$100\,mK with the help of a capillary-fed superfluid tank and a closed cycle dilution refrigerator. The main tank is supported with fiberglass flexures and is encased in two layers of vapor-cooled shields which allow Taurus to make full use of the extended flight time offered by the super-pressure balloon platform. The Taurus cryostat is projected to hold for over 50 days while weighing under 1000\,lbs.  We present the design, testing, and thermal analysis of the Taurus cryogenic systems.
\end{abstract}


\section{INTRODUCTION}
\label{sec:intro}  % \label{} allows reference to this section

The Cosmic Microwave Background (CMB) has long been a rich resource of cosmological information. State-of-the-art measurements of the CMB still find large angular scales to be challenging, with current ones not being cosmic variance limited. Taurus is a balloon-based CMB telescope that aims to map $\sim$70\% of the sky with high fidelity at large angular scales ($\ell < 30$). An instrument overview is presented by May et al.~2024\cite{jared}.

Taurus is targeting a 2027 flight on a NASA super-pressure balloon (SPB), and the instrument will employ 10\,000 transition-edge bolometers cooled to $\leq$100\,mK and split into four frequency bands centered around 150, 220, 280, and 350\,GHz. Taurus is designed with large angular scale systematic mitigations in mind -- employing de-pointed receivers and a scan strategy which modulates the sky past typical detector 1/f knees. The reduced loading environment offered by the scientific balloon platform will allow Taurus to improve measurements of the optical depth to reionization ($\tau$) and produce high fidelity polarized Galactic dust maps.

To maximize its scientific return, Taurus should have a cryogenic hold time $>$50 days and thus requires a well-optimized cryostat. The mass and power constraints set by the balloon platform impose a liquid helium only cryostat with vapor-cooled radiation shields. These cryostat architectures have been successfully demonstrated on the scientific balloon platform and provide the best hold time to mass ratio while mitigating complexity \cite{spider_cryo,blasttng}.

The Taurus cryostat is a 660\,L liquid helium only cryostat with two vapor-cooled radiation shields designed to hold for 50 days. The mechanical design is shown in figure \ref{fig:solid_render} and the thermal layout is presented in figure \ref{fig:flow_diagram}. A 1.5\,K thermal point is created by a capillary-fed helium-4 super-fluid tank which is evacuated to the ambient pressure at the float altitude of the balloon. Sub-K cooling is provided by a set of closed-cycle dilution refrigerators backed with helium-3 absorption fridges. The thermal architecture above 4\,K is very similar to the \textsc{Spider} cryostat which has flown two successful missions on the Long Duration Balloon platform\cite{spider_architecture}. However, unlike \textsc{Spider} and other predecessors, Taurus will fly on a Super-pressure balloon which should provide a longer flight time at the expense of a smaller mass budget. Thus, the Taurus cryostat has to be lighter yet hold for a longer period\cite{nasa_balloon}.

\section{CRYOSTAT DEVELOPMENT}
\subsection{Cryogenic Layout}\label{sec:layout}

The Taurus flight cryostat must be designed to maximize hold time while remaining within the mass budget allocated by the super-pressure balloon platform. We want to pack as large a main tank (MT) as the mass constraint allows while minimizing the ratio of surface area to volume to reduce radiative coupling between stages. Therefore, the optimal cryogenic layout would minimize the surface area of the 4~K shield which encapsulates the three cylindrical telescopes while obeying the de-pointed constraint, (see sec. \ref{sec:intro}). Multiple families of solutions were investigated but it was found that around the optimum a broad variety of solution results in similar surface areas. Ultimately, the cryogenic layout was chosen to optimize for serviceability and manufacturability while remaining within 10\% of optimal.

The resulting optics layout is placing the two larger low-frequency telescopes next to each other at 35$^\circ$ elevation and opposing azimuth. The third smaller high-frequency telescope is placed in the V formed between the two also at 35$^\circ$ elevation.  A render of the cryostat is presented in figure \ref{fig:solid_render} showing how the optics trusses and receivers are positioned.

\newlength{\imageheight}
\settoheight{\imageheight}{\includegraphics[width=0.45\textwidth]{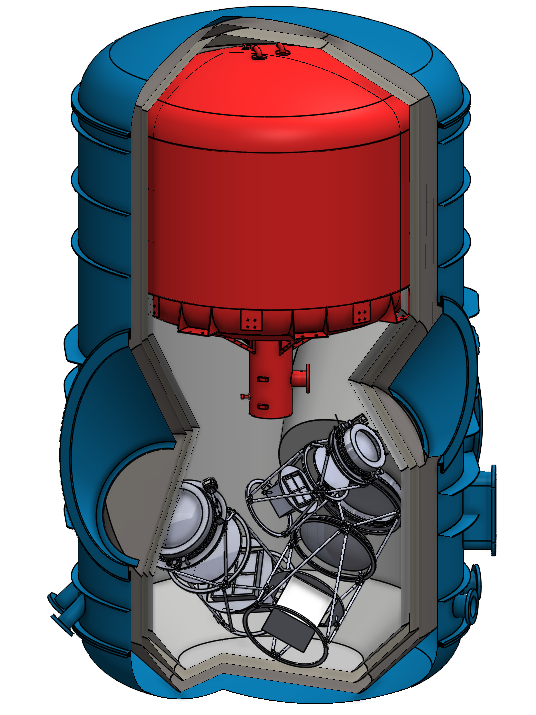}}
\newlength{\padding}
\setlength{\padding}{0.5cm} % Adjust padding as needed

\begin{figure}[ht]
    \centering
    \begin{minipage}[b]{0.45\textwidth}
        \includegraphics[width=\textwidth]{figures/cryostat_full_4.PNG}
        \caption*{}
    \end{minipage}
    \begin{minipage}[b]{0.05\textheight}
        % \vspace*{\fill}
        \centering
        \begin{tikzpicture}
            % Draw the vertical dimension line with padding
            \draw[thick] (0,0) -- (0,\imageheight-\padding);
            % Draw horizontal bars at the ends
            \draw[thick] (-0.2,0) -- (0.2,0);
            \draw[thick] (-0.2,\imageheight-\padding) -- (0.2,\imageheight-\padding);
            % Add the vertical text with offset
            \node[anchor=center, rotate=90, yshift=0.3cm] at (0,\imageheight/2) {\Large{2.5\,m}};
        \end{tikzpicture}
        \vspace*{\padding}
    \end{minipage}
    \begin{minipage}[b]{0.45\textwidth}
        \includegraphics[width=\textwidth]{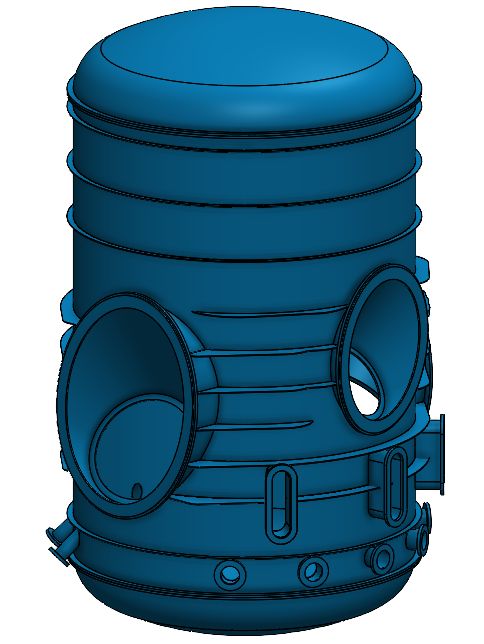}
        \caption*{}
    \end{minipage}
    
    \caption{Simplified Taurus cryostat renders. (left) Cutaway showing the receiver layout and internal layers. The main tank (MT -- red), 4K shroud (light grey), vapor cooled shields (VCS1\&2 -- grey and dark grey), and vacuum vessel (VV -- blue). (right) Vacuum vessel (VV) render showing penetrations and ribbing needed to stiffen the shell. The vacuum vessel is 2.5\,m tall and 1.2\,m in diameter.}
    \label{fig:solid_render}
\end{figure}

\subsection{Mass Optimization}

To optimize the mass of a system such as Taurus, the first step is to try and minimize its overall size by selecting the appropriate layout (see sec \ref{sec:layout}). However, simply choosing the smallest surface area may not result in the lightest cryostat. The cryostat's shells are subject to various pressure loads and thus must be made thicker if their geometry does not respond to the strain favorably.  Often a larger curvy-simple shape (such as a cylinder or dome) will end up being a less massive solution than the optimal complex geometry. Taurus follows in this philosophy by having a cylindrical main tank with torispherical domes on either side and a similarly shaped vacuum vessel with the minimal required complexities to accommodate the instrument's optical windows.

Material choices can also greatly influence the mass of the cryogenic system and thus designers tend to specify materials with the highest possible yield strength-to-weight ratio. A common choice for cryostats is 5083 Aluminium, however when welded, this material returns to the annealed state and loses the yield strength to weight ratio advantage\cite{alu_postweld}. Taurus chose to use an AISI 304L main tank as it maintains post-welding strength and is much easier to weld which ideally will result in no cryogenic leaks\cite{SS_postweld}. Sticking to the same material for plumbing and the MT reduces the number of bi-metal junctions in the cryostat which are prone to developing leaks. The vacuum vessel for Taurus will nevertheless be made out of 5083 aluminium, even though it anneals after welding. For shells subject to external pressures, less dense but thicker shells (such as aluminium) reduce the amount of ribbing needed to prevent buckling. The safety factors for the Taurus cryostat are around 1.5 so we will measure the physical properties of the formed and welded material to confirm that they exceed those assumed during the design.

\subsection{Flexures}

Balloon payloads are subject to large mechanical stresses during launch and recovery. In addition to absolute strength, the connection between outer and inner stages needs to be as stiff as possible to minimize deflections and push resonant frequencies out of the detector bandwidth. Therefore, connecting the inner temperature stages to the vacuum vessel presents mechanical and thermal requirements which are at odds, and must balance the admitted heat leak with the mechanical strength. Finally, cryogenic flexures must be able to accommodate the differential thermal contractions, which are several millimeters for a cryostat the size of Taurus's.

\begin{figure}
    \centering
\includegraphics[width=0.45\textwidth]{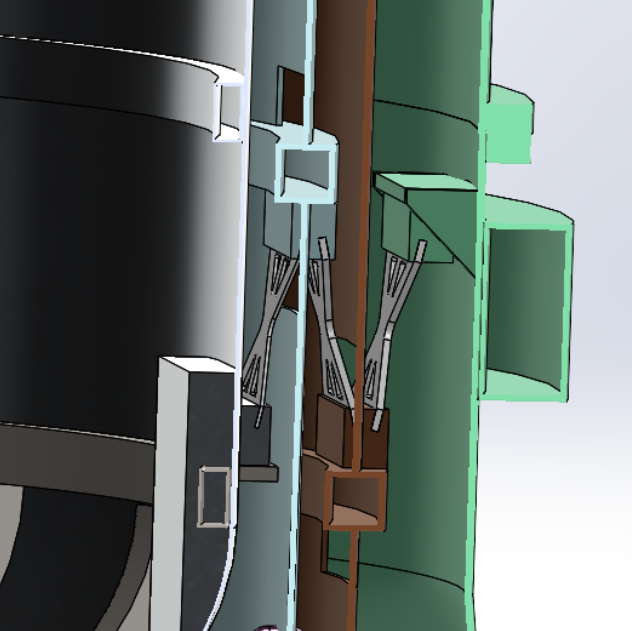}
\includegraphics[width=0.45\textwidth]{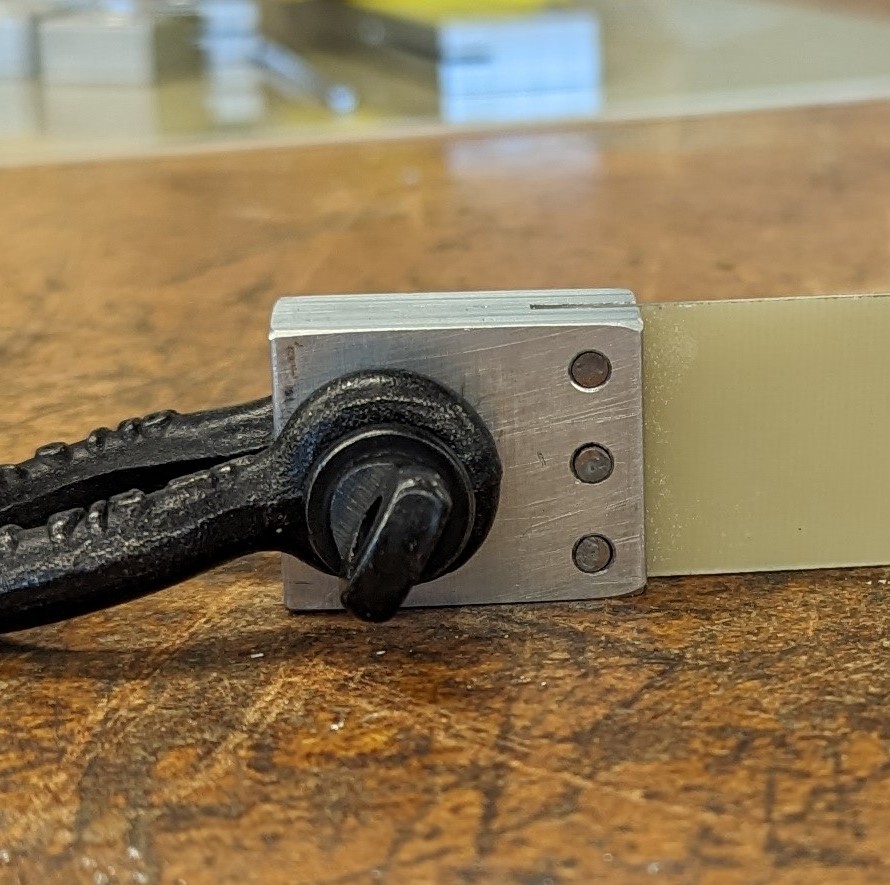}
\vspace{0.2cm}
    \caption{(left) Cut away showing one of the members for the 8 sets of flexures running from the VV to the MT via the two VCSs. The shape was optimized to minimize the heat-leak while maintaining adequate strength and stifness. (right) Test coupon used to validate the G10-aluminium interface which Taurus will use.}
    \label{fig:flexure_picture}
\end{figure}

The proposed design for Taurus is similar to the designs used in other large cryostats such as \textsc{Spider}\cite{spider_architecture} and SO-LAT\cite{SO_LAT_overview}. The cryostat is supported internally by a partial Vierendeel truss made from G10-CR plates near the center of mass of the system. There are three sets of flexures running sequentially between the temperature stages, each of which is made up of eight individual G10-CR plates, a cut-away of one such set is shown in figure \ref{fig:flexure_picture}. Staggering the flexures between temperature stages allows for the VCSs to remain fully light tight without requiring bespoke tape-joints and mitigates the risk of large light leaks. An as-built flexure which was mechanically tested to have sufficient strength and stiffness has been subjected to a cryogenic thermal conductivity test to validate it against the thermal model \ref{sec:thermal_model}. These tests show that under steady-state the flexures admit for 50\,mW of heat-leak into the 4.2\,K main tank.

A scale version of the Taurus flexure truss has been built and tested to validate the design and simulations used for this critical element of the cryostat. The resonant frequencies and buckling loads have been measured both at room and cryogenic temperatures in preparation for the full scale flexure commissioning.

G10 to aluminium transitions can be difficult to design and may fail after thermal and mechanical cycles. Some experiments have solved the problem by avoiding the use of glue entirely (SO-SAT\cite{SO_SAT}) while others do extensive testing on engineered glue joints. Taurus uses a combination of both: an engineered glue joint backed-up with mechanically constraining pins as seen in figure \ref{fig:flexure_picture}. These joints have been built and extensively tested to demonstrate that both elements are capable of sustaining working loads individually such that if either fails the other will be sufficient. The results for a subset of performed tests are presented in table \ref{tab:pull_tests}.

\begin{table}
\centering
\caption{Summary of tests done on a 1"x1/16" G10-CR sample placed in a 3/8" slot. Reduced strength is presented per G10 cross-section for pined joints and per glue cross-section for glued joints. The pull tester used maxed out at 2000\,lbs resulting in lower limits for glued joints. Samples that were thermally cycled and pulled cold were dipped in liquid nitrogen. Scaling the tested samples to full-sized Taurus flexures predicts that the joint strength always exceeds the base material failure load.}
\vspace{0.2cm}
\begin{tabular}{|>{\raggedright\arraybackslash}m{6cm}|>{\centering\arraybackslash}m{3cm}|>{\centering\arraybackslash}m{3.5cm}|}
\hline
\textbf{Sample} & \textbf{Joint strength} & \textbf{Reduced strength} \\
\hline
3x1/8"~OD pins & 750\,lbs/in & 12\,ksi \\
\hdashline
EA E-120HP with 0.005" microspheres, pulled cold after 20 thermal cycles & $>$2000\,lbs/in & $>$2666\,psi \\
\hdashline
Pins and glue (combined above 2 rows), pulled cold after 20 thermal cycles & $>$2000\,lbs/in & \makecell[tc]{Matches above, \\ Glue did not fail} \\
\hdashline
9309.3NA glue, pulled cold after 10 thermal cycles & $>$2000\,lbs/in & $>$2666\,psi \\
\hline
\end{tabular}
\vspace{0.2cm}
\label{tab:pull_tests}
\end{table}

\subsection{Sub-Kelvin systems}

To reach the 100\,mK base temperature Taurus will use set of cascaded coolers backed by the large 4.2\,K thermal bath provided by the evaporating helium in the MT. The first such cooler is 5\,L super-fluid tank (SFT) which is connected to the MT with a set of capillaries and pumped down to roughly 7\,torr by the ambient pressure at float. Capillary fed SFTs have a lot of cryogenic heritage but come with the downside that the flow in the capillaries can not be tuned after commissioning\cite{spider_cryo, sft_seminal}. This can be resolve by running multiple capillaries in parallel with a small resistive heating element at the midpoint of each. Since super-fluid helium has no viscosity and a very high effective thermal conductivity the majority of the flow impedance comes from the small fraction of normal helium on the MT side of the capillary tube. By heating the midpoint past the superfluid transition temperature, the normal fluid fraction, and thus impedance, raises drastically reducing the mass flow rate\cite{sft_seminal}. 

The final stage of refrigeration is a closed-cycle helium3-4 dilution refrigerator (DR) developed by Chase Research Cryogenics with a heat lift of 3\,$\mu$W at 100\,mK\cite{chase_miniDR}. The DR circulates mixture using a condensing pump which must be cooled to 300\,mK.  Taurus does not require continuous cooling to 100\,mK as it only observes at night, so we plan to cool the DR's condensing pump with a set of \textsc{Spider} helium-3 absorption fridges in parallel backed by the SFT. Cycling the fridges shortly before sunset should provide a stable 100\,mK for the duration of the observing night and allow Taurus to benefit from the extremely stable loading environment at float.

\section{Thermal Modeling}\label{sec:thermal_model}

To validate and inform the design of the cryostat a thermal model was developed. This thermal model is a lumped element quasi-static solver which models interactions between isothermal stages. For modeling purposes, that the cryostat has two fixed temperature boundary conditions: the vacuum vessel at 300\,K and the main tank at 4.2\,K. Solutions for heat flow and equilibrium temperatures can then be obtained in steady state.

\begin{figure}
    \centering
    \includegraphics[width=0.95\textwidth,trim={0 3.7cm 0 0},clip]{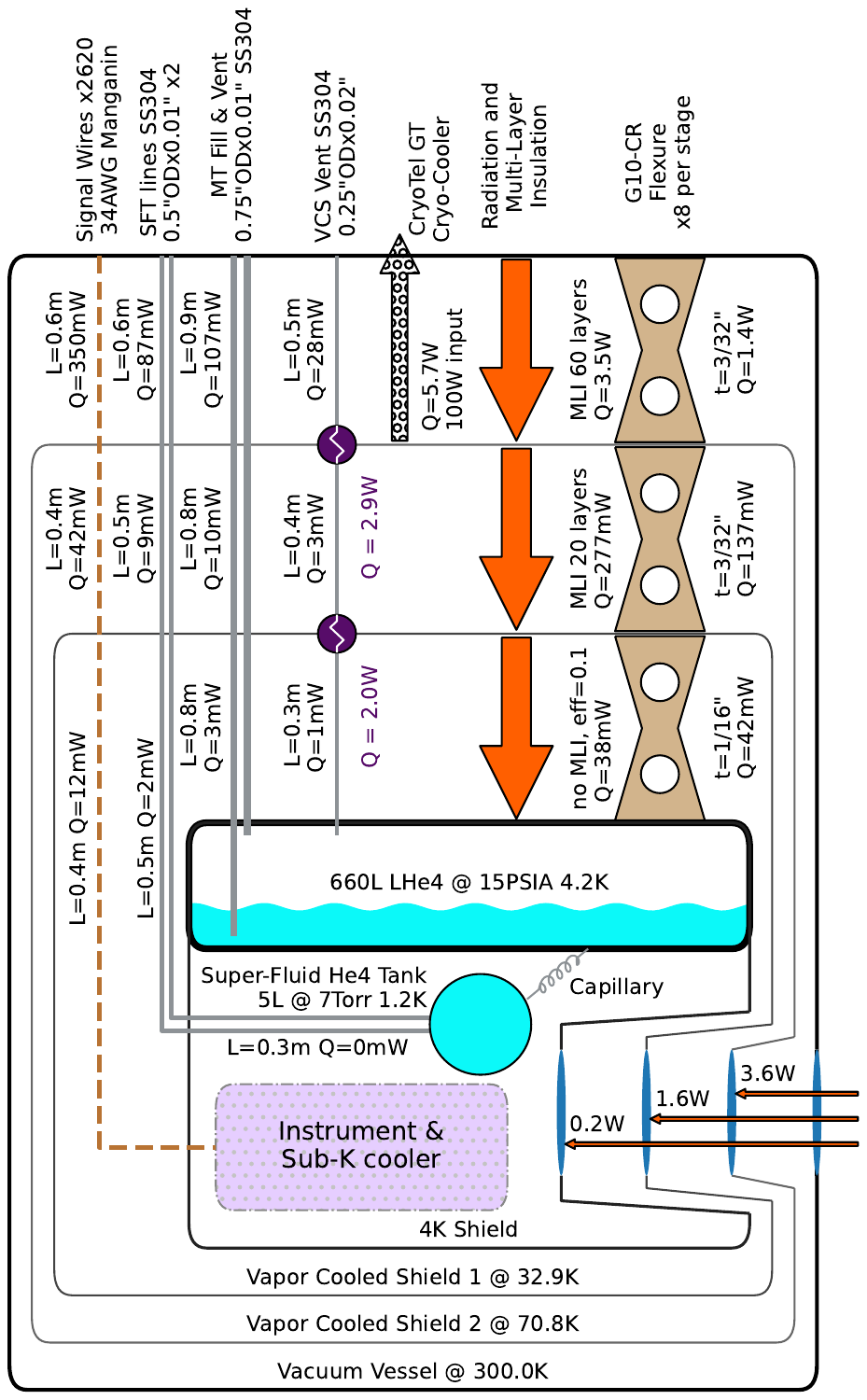}
    \caption{Summary of heat exchanges in the Taurus flight cryostat. The case shown here corresponds to ``Cryo-Cooler (100\,W input)" in table \ref{tab:thermal_cases}. Sub-K stages and instrument are not included in the cryostat thermal model and thus not shown.}
    \label{fig:flow_diagram}
\end{figure}

The results of this model are presented in figure \ref{fig:flow_diagram}. Interactions between each modeled temperature stage are shown, with resulting heat flow and final temperatures. 

\subsection{IR Optical Filters}

The Taurus cryostat will have three optical windows which must be transparent to mm-wave radiation while filtering out IR to limit radiative coupling between stages. Such low-pass optical filters can be challenging to model. Ultimately, measured values from SPIDER were used after being scaled to Taurus' aperture sizes\cite{spider_cryo}.

Assuming that the empirically estimated 200\,mW of power is deposited on the main tank by the windows, the sum total of all other inputs is around 210\,mW.  Optimizing further will result in diminishing returns as the window load will become massively dominant. In fact, for all temperature stages this poorly understood input dominates loading. 

Since the model is dominated by an empirically measured quantity, inaccuracies in the rest of the model are suppressed. In fact, almost no thermal modeling could have been done to size the main tank - assuming that the parasitic heat load on the MT would have similar magnitude to the window load and then choosing a volume of helium that would last 50 days at that input power would have given a reasonably accurate estimate for the required main tank size. The thermal model only serves as a design tool which helps us make decisions to make sure that parasitic heat-leak through other elements are sub-dominant to the unavoidable heat leak generated by the windows.

\subsection{Multi Layer Insulation}

We plan to use multi-layer insulation (MLI) blankets between all temperature stages of the Taurus cryostat. In a purely radiative environment, $N$ layers of MLI will reduce the transferred power by a factor of $1/N$. This ideal performance is nearly attainable with MLI operating between room temperature hot sides $T_h$ and cryogenic cold sides $T_c$. However, as $T_h$ is reduced, conduction between layers becomes non-negligible compared to radiative coupling and thus the insulation's performance deviates from the ideal model. In fact, with $T_h$ dropping under 100\,K MLI blankets can become conductive shorts between temperature stages and increase thermal flux as compared to not having any blanket at all\cite{keller1974thermal,mli_cryogenic}.

The interface between the vacuum vessel and VCS2 is dominated by radiation and thus MLI will have the largest effect; we plan to use a 60 layer blanket here. For the VCS1--VCS2 interface, which is expected to be around 50\,K to 110\,K, we plan to use 20 very loosely packed layers which are physically attached to VCS1. Anchoring the blanket to the cold side limits the possible harm done by the blanket by preventing it from creating a thermal conduction path between the two shells. Similarly, for the MT-VCS1 interface we will also place around 10 layers mostly to mitigate light leaks and decrease the emissivity of the bare surfaces.

For modeling purposes the VV--VCS2 and VCS2--VCS1 MLI is modeled with Lockheed MLI equations 4-11 as the data used to create this equation includes $T_c$ temperatures as low as 39\,K and $T_h$ temperatures above 300\,K\cite{keller1974thermal}. MT--VCS1 is modeled as having the effective emissivity of aluminized mylar but with no MLI. The literature is filled with a broad range of as-built MLI conductance measurements and thus as-built performance is likely to vary by as much as 5 times\cite{mli_cryogenic,keller1974thermal}.

\subsection{Hold-Time Optimization}\label{sec:holdtime}

A few options were explored to extend the hold time of the Taurus flight cryostat during the design phase. Three cases that extend the hold time by around 20\% are presented, in addition to the base case, in table \ref{tab:thermal_cases}. The presented case are: including a Stirling cooler to super cool VCS2, increasing the size of the main tank to match the performance gained by the cryo-cooler and finally adding a third VCS around VCS2 to maximize the extracted enthalpy from the escaping helium vapor.  As a figure of merit the mass of each solution was estimated.

 % Adjust row height (1.5 times the default)

\begin{table}
    \centering
    \caption{Comparison of explored schemes which can be used to increase the hold time of the Taurus flight cryostat. Added mass is computed by adjusting the solid models to get an accurate estimate. The cryo-cooler case will be adopted by the Taurus cryostat due to its small mass penalty and ability to be descoped as described in section \ref{sec:holdtime}.}
    \vspace{0.2cm}
    \begin{tabular}{|m{3cm}|>{\centering\arraybackslash}m{4cm}|>{\centering\arraybackslash}m{3cm}|>{\centering\arraybackslash}m{4cm}|}
        \hline
        \textbf{Case} & \textbf{Vapor Cooled Shield Temperature (K)} & \textbf{Hold Time (days)} & \textbf{Added Mass (estimated, lbs)} \\
        \hline
        Base & 42.0 -- 112.3 & 48.4 & +0 \\
        \hdashline
        \textbf{Stirling-Cooler (100\,W input)} & \textbf{34.5 -- 69.8} & \textbf{59.8} & \textbf{+55 (battery \& solar)} \\
        \hdashline
        Larger MT &  42.1 -- 112.0 & 59.7 & +60 (scale cryostat) \\
        \hdashline
        Extra VCS & 35.0 -- 72.0 -- 154.3  & 59.7 & +75 (larger VV \& shield) \\
        \hline
    \end{tabular}
    \label{tab:thermal_cases}
\end{table}

Ultimately we plan to include all the features needed to use a Stirling cryo-cooler on the cryostat (a vacuum flange and provisions for the extra batteries)\cite{cooler}. It is a particularly attractive option for the following reasons:

\begin{enumerate}
    \item It is descopable: if upon delivery of the cryostat the hold time is adequate without a cooler we can choose to not fly one. All other options need to be done during commissioning.
    \item The load curve of a Stirling cryo-cooler is very flat around 80\,K thus even with a large uncertainty on the window and MLI loading on VCS2 a cryo-cooler can guarantee a VCS2 temperature around 80\,K. Having a fixed boundary condition at 80\,K makes the rest of the thermal model much more believable.
\end{enumerate}

Upon receiving the cryostat we will measure the as-built hold time and be able to tune the cryo-cooler input power to achieve the desired hold time, possibly even removing it entirely if the cryostat performs nominally.

\section{Conclusion}

Taurus aims to map the CMB with high fidelity at large angular scales. To achieve these goals Taurus is built with low $l$ systematic mitigations such as de-pointed receivers and a rapid scan which modulates the sky. The combination of these science requirements and the constraints set by the super-pressure scientific ballooning platform result in a challenging design problem for the Taurus flight cryostat. 

We present the mechanical design and thermal architecture of the Taurus cryogenic systems. The cryostat developed for Taurus is optimized both for mass and hold time. The cryostat weighs under 1000\,lbs and is projected to remain cryogenic for over 50 days. The flexure design is discussed in detail and we present qualifying tests that demonstrate the mechanical strength of the G10 and aluminium composite structure. Finally, the cryostat's thermal architecture is validated with a thermal model that predicts a 59.8 day hold-time when supplemented with a Stirling-cooler.

With the presented thermal and mechanical design, Taurus aims to improve measurements of the optical depth to reionization ($\tau$) and produce high-fidelity polarized Galactic dust maps.

\appendix    %>>>> this command starts appendixes

\acknowledgments % equivalent to \section*{ACKNOWLEDGMENTS}       
 
Taurus is supported in the USA by NASA award number 80NSSC21K1957. JEG and TJLJG acknowledge support from The Icelandic Research Fund (Grant number: 2410656-051) and the European Union (ERC, CMBeam, 101040169).

% References
\bibliography{report} % bibliography data in report.bib

\begin{thebibliography}{10}

\bibitem{jared}
May, J. et~al., ``Instrument overview of taurus: A balloon-borne cmb and dust polarization experiment,'' {\em Society of Photo-Optical Instrumentation Engineers (SPIE) Conference Series}  (2024).

\bibitem{spider_cryo}
Gudmundsson, J., Ade, P., Amiri, M., Benton, S., Bock, J., Bond, J., Bryan, S., Chiang, H., Contaldi, C., Crill, B., Dore, O., Filippini, J., Fraisse, A., Gambrel, A., Gandilo, N., Hasselfield, M., Halpern, M., Hilton, G., Holmes, W., Hristov, V., Irwin, K., Jones, W., Kermish, Z., MacTavish, C., Mason, P., Megerian, K., Moncelsi, L., Montroy, T., Morford, T., Nagy, J., Netterfield, C., Rahlin, A., Reintsema, C., Ruhl, J., Runyan, M., Shariff, J., Soler, J., Trangsrud, A., Tucker, C., Tucker, R., Turner, A., Wiebe, D., and Young, E., ``The thermal design, characterization, and performance of the s pider long-duration balloon cryostat,'' {\em Cryogenics}~{\bf 72},  65–76 (Dec. 2015).

\bibitem{blasttng}
Galitzki, N., Ade, P., Angilè, F.~E., Ashton, P., Austermann, J., Billings, T., Che, G., Cho, H.-M., Davis, K., Devlin, M., Dicker, S., Dober, B.~J., Fissel, L.~M., Fukui, Y., Gao, J., Gordon, S., Groppi, C.~E., Hillbrand, S., Hilton, G.~C., Hubmayr, J., Irwin, K.~D., Klein, J., Li, D., Li, Z.-Y., Lourie, N.~P., Lowe, I., Mani, H., Martin, P.~G., Mauskopf, P., McKenney, C., Nati, F., Novak, G., Pascale, E., Pisano, G., Santos, F.~P., Scott, D., Sinclair, A., Soler, J.~D., Tucker, C., Underhill, M., Vissers, M., and Williams, P., ``Instrumental performance and results from testing of the blast-tng receiver, submillimeter optics, and mkid detector arrays,'' in [{\em Millimeter, Submillimeter, and Far-Infrared Detectors and Instrumentation for Astronomy VIII}{\nolinebreak\hspace{0.1em}]},  Holland, W.~S. and Zmuidzinas, J., eds., SPIE (July 2016).

\bibitem{spider_architecture}
Gudmundsson, J.~E., Ade, P. A.~R., Amiri, M., Benton, S.~J., Bihary, R., Bock, J.~J., Bond, J.~R., Bonetti, J.~A., Bryan, S.~A., Burger, B., Chiang, H.~C., Contaldi, C.~R., Crill, B.~P., Doré, O., Farhang, M., Filippini, J., Fissel, L.~M., Gandilo, N.~N., Golwala, S.~R., Halpern, M., Hasselfield, M., Hilton, G., Holmes, W., Hristov, V.~V., Irwin, K.~D., Jones, W.~C., Kuo, C.~L., MacTavish, C.~J., Mason, P.~V., Montroy, T.~E., Morford, T.~A., Netterfield, C.~B., O’Dea, D.~T., Rahlin, A.~S., Reintsema, C.~D., Ruhl, J.~E., Runyan, M.~C., Schenker, M.~A., Shariff, J.~A., Soler, J.~D., Trangsrud, A., Tucker, C., Tucker, R.~S., and Turner, A.~D., ``Thermal architecture for the spider flight cryostat,'' in [{\em Millimeter, Submillimeter, and Far-Infrared Detectors and Instrumentation for Astronomy V}{\nolinebreak\hspace{0.1em}]},  Holland, W.~S. and Zmuidzinas, J., eds., SPIE (July 2010).

\bibitem{nasa_balloon}
NASA, ``Suborbital research.'' \url{https://science.nasa.gov/researchers/suborbital}.
\newblock Accessed in June 2024.

\bibitem{alu_postweld}
Collette, M., ``The impact of fusion welds on the ultimate strength of aluminum structures,'' {\em 10th International Symposium on Practical Design of Ships and other Floating Structures, PRADS 2007}~{\bf 2} (01 2007).

\bibitem{SS_postweld}
Kellai, A., Kahla, S., Dehimi, S., Kaba, L., and Boutaghou, Z., ``Effect of post weld heat treatment on the microstructure and mechanical properties of a gas-tungsten-arc-welded 304 stainless steel,'' {\em The International Journal of Advanced Manufacturing Technology}~{\bf 121} (08 2022).

\bibitem{SO_LAT_overview}
Zhu, N., Orlowski-Scherer, J.~L., Xu, Z., Ali, A., Arnold, K.~S., Ashton, P.~C., Coppi, G., Devlin, M.~J., Dicker, S., Galitzki, N., Gallardo, P.~A., Henderson, S.~W., Ho, S.-P.~P., Hubmayr, J., Keating, B., Lee, A.~T., Limon, M., Lungu, M., Mauskopf, P.~D., May, A.~J., McMahon, J., Niemack, M.~D., Piccirillo, L., Puglisi, G., Rao, M.~S., Salatino, M., Silva-Feaver, M., Simon, S.~M., Staggs, S., Thornton, R., Ullom, J.~N., Vavagiakis, E.~M., Westbrook, B., and Wollack, E.~J., ``Simons observatory large aperture telescope receiver design overview,'' (2018).

\bibitem{SO_SAT}
Galitzki, N., Tsan, T., Spisak, J., Randall, M., Silva-Feaver, M., Seibert, J., Lashner, J., Adachi, S., Adkins, S.~M., Alford, T., Arnold, K., Ashton, P.~C., Austermann, J.~E., Baccigalupi, C., Bazarko, A., Beall, J.~A., Bhimani, S., Bixler, B., Coppi, G., Corbett, L., Crowley, K.~D., Crowley, K.~T., Day-Weiss, S., Dicker, S., Dow, P.~N., Duell, C.~J., Duff, S.~M., Gerras, R.~G., Groh, J.~C., Gudmundsson, J.~E., Harrington, K., Hasegawa, M., Healy, E., Henderson, S.~W., Hubmayr, J., Iuliano, J., Johnson, B.~R., Keating, B., Keller, B., Kiuchi, K., Kofman, A.~M., Koopman, B.~J., Kusaka, A., Lee, A.~T., Lew, R.~A., Lin, L.~T., Link, M.~J., Lucas, T.~J., Lungu, M., Mangu, A., McMahon, J.~J., Miller, A.~D., Moore, J.~E., Morshed, M., Nakata, H., Nati, F., Newburgh, L.~B., Nguyen, D.~V., Niemack, M.~D., Page, L.~A., Sakaguri, K., Sakurai, Y., Rao, M.~S., Saunders, L.~J., Shroyer, J.~E., Sugiyama, J., Tajima, O., Takeuchi, A., Bua, R.~T., Teply, G., Terasaki, T., Ullom, J.~N., Lanen, J. L.~V., Vavagiakis, E.~M.,
  Vissers, M.~R., Walters, L., Wang, Y., Xu, Z., Yamada, K., and Zheng, K., ``The simons observatory: Design, integration, and testing of the small aperture telescopes,'' (2024).

\bibitem{sft_seminal}
Fujiyoshi, Y., Mizusaki, T., Morikawa, K., Yamagishi, H., Aoki, Y., Kihara, H., and Harada, Y., ``Development of a superfluid helium stage for high-resolution electron microscopy,'' {\em Ultramicroscopy}~{\bf 38}(3),  241--251 (1991).

\bibitem{chase_miniDR}
Teleberg, G., Chase, S., and Piccirillo, L., ``A cryogen-free miniature dilution refrigerator for low-temperature detector applications,'' {\em Journal of Low Temperature Physics}~{\bf 151},  669--674 (01 2008).

\bibitem{keller1974thermal}
Keller, C., Cunnington, G., and Glassford, A., ``Thermal performance of multilayer insulations,'' tech. rep. (1974).

\bibitem{mli_cryogenic}
Ross, R., ``Quantifying mli thermal conduction in cryogenic applications from experimental data,'' in [{\em IOP Conference Series: Materials Science and Engineering}{\nolinebreak\hspace{0.1em}]},   {\bf 101}(1),  012017, IOP Publishing (2015).

\bibitem{cooler}
Sunpower, ``Cryotel gt.'' \url{https://www.sunpowerinc.com/products/stirling-cryocoolers/cryotel-cryocoolers/gt}.
\newblock Accessed in June 2024.

\end{thebibliography}
\bibliographystyle{spiebib} % makes bibtex use spiebib.bst

\end{document}